# A Computational Study of Hafnia based Ferroelectric Memories: From Ab-initio via Physical Modeling to Circuit Models of Ferroelectric Device

Milan Pešić[1], Christopher Künneth[2], Michael Hoffmann[1], Halid Mulaosmanovic[1], Stefan Müller[3], Evelyn T. Breyer[1], Uwe Schroeder[1], Alfred Kersch[2], Thomas Mikolajick[1,4], Stefan Slesazeck[1]

[1]*NaMLab gGmbH, Noethnitzer Str. 64, Dresden D-01187, Germany*
[2]*Department of Applied Sciences and Mechatronics, Munich University of Applied Sciences, Munich, Germany*
[3]*Ferroelectric Memory GmbH, Dresden D-01187, Germany*
[4]*Chair of Nanoelectronic Materials, Institute of Semiconductors and Microsystems, TU Dresden*
Milan.Pesic@namlab.com

**Abstract.** The discovery of ferroelectric properties of binary oxides revitalized the interest in ferroelectrics and bridged the scaling gap between the state-of-the-art semiconductor technology and ferroelectric memories. However, before hitting the markets, the origin of ferroelectricity and in-depth studies of device characteristics are needed. Establishing a correlation between the performance of the device and underlying physical mechanisms is the first step toward understanding the device and engineering guidelines for a novel, superior device. Therefore, in this paper a holistic modeling approaches which lead to a better understanding of ferroelectric memories based on hafnium and zirconium oxide is addressed. Starting from describing the stabilization of the ferroelectric phase within the binary oxides via physical modeling the physical mechanisms of the ferroelectric devices are reviewed. Besides, limitations and modeling of the multi-level operation and switching kinetics of ultimately scaled devices as well as the necessity for Landau-Khalatnikov approach are discussed. Furthermore, a device level model of ferroelectric memory devices that can be used to study the array implementation and their operational schemes are addressed. Finally, a circuit model of the ferroelectric memory device is presented and potential further applications of ferroelectric devices are outlined.

## 1 Introduction

In the age of mobile devices nonvolatile memories have become an essential part of every electronic system. Traditionally nonvolatile memories are limited with respect to re-write speed and cycling endurance due to the so-called "voltage time dilemma" [1]. Additionally the high voltages required during the re-write process make their integration into scaled complementary metal-oxide-semiconductor (CMOS) processes more and more complicated. The stable remanent polarization at zero electric field together with the fact that the polarization can be switched by an electrical field makes ferroelectrics a natural choice to realize a nonvolatile memory [2, 3]. Therefore, already in the 1950s the concept of a ferroelectric memory was proposed [4] and first attempts were made to realize solid state memories based on the ferroelectric Bariumtitanate [5, 6]. However, these attempts were based on pure cross point arrays and the resulting disturb issues could not be solved. The rapid development of semiconductor technology enabled to solve this fundamental issue by adding a MOS select transistor to the cell architecture. As a result the first commercial ferroelectric memories appeared in the early 1990s [7]. This success spurred the hope that within a few years a high density memory that would have the cell size and performance of DRAM and at the same time be nonvolatile would emerge [8]. Soon, however, it was recognized that the integration of very complicated perovskite or even layered perovskite materials requires a large integration effort that would hinder a scaling on the same pace as CMOS and related memory concepts based on floating gates or charge trapping layers [9]. Since the readout of a ferroelectric capacitor is based on the charge that is transferred during switching [1] a 3-dimensional capacitor would be required below 100 nm groundrules. This is adding new integration challenges for the complex materials not only of the ferroelectric itself but also the required metal-oxide electrodes. For this challenge no solution can be provided yet [10, 11]. Therefore, the development of such perovskite based memories were successful but at the same time seem to have hit a wall at about the 100 nm groundrules [12]. Another path to scale the memory cell without going into the third dimension is the integration of the functional ferroelectric film into a field effect transistor. Such an approach, called the ferroelectric field effect transistor (FeFET) was proposed already in the late 1950s [13, 14]. However, perovskite or layered perovskite ferroelectrics like lead-zirconate-titanate (PZT) and strontium-bismuth-tantalate (SBT) in contact with silicon require an interface buffer layer to avoid silicidation reactions. Due to the high permittivity of perovskites it is nearly impossible to match the permittivity of the ferroelectric with a non-ferroelectric interface layer while trying to reduce the gate stack thickness. An internal depolarization field under zero applied voltage is the consequence leading to poor data retention [15]. Using hafnium oxide as the interface layer between a very thick SBT layer and silicon [16], nonvolatile retention could be demonstrated. However, the thickness requirements for the SBT layer hinder the scaling of such devices to the deep sub-100 nm regime as well.

The discovery of ferroelectricity in hafnium oxide by Boescke et al. in 2011 [17] may resolve these historic issues of ferroelectric memory devices explained above. Hafnium oxide is the standard gate dielectric in sub 45 nm CMOS processes. Proven fabrication and integration schemes exist and hafnium oxide is thermodynamically stable on silicon [18]. Therefore this discovery has drastically changed the view on ferroelectric memories [19]. Both the scaling of classical capacitor based memories as well as the realization of FeFET based memories seem to be in reach again [20, 21] and even the possibility of making classical DRAM non-volatile has been pointed out recently [22, 23]. However, in order to apply the unexpected ferroelectric behavior of hafnium oxide to reliable products a number of fundamental questions need to be answered: the origin of the ferroelectric phase, the control of the influencing process parameters, the device and array concepts as well as the degradation under use conditions. With such fundamental questions to be addressed, the modelling on all abstraction levels from atomistic material simulations via device simulations up to the array and circuit level are required. This paper summarizes the recent progress on this exciting topic.

## 2 Engineering the Material: *Ab-initio* Study

The observation of ferroelectricity in thin hafnia films has been surprising since the most stable monoclinic $P2_1/c$ phase, the high temperature tetragonal $P4_2/nmc$ and cubic $Fm3m$ phases and the high pressure orthorhombic $Pbca$ phase (the well-known equilibrium phases) are centrosymmetric and non-polar. Following older experimental [24] and theoretical [25] work, the metastable, polar, orthorhombic $Pca2_1$ was proposed as the ferroelectric phase, which is nearly indistinguishable from tetragonal in grazing incidence X-ray diffraction (GIXRD). With a combined transmission electron microscopy (TEM) and nanoscale electron diffraction study, Sang et al. [26] were able to identify the $Pca2_1$ phase in a thin, ferroelectric hafnia film with some uncertainty with regard to the $Pbca$ phase.

Density functional theory (DFT) total energy calculations qualitatively mostly agree that the $Pca2_1$ phase of hafnia is the second most stable phase with about 60 meV/f.u., followed by the tetragonal phase with 95 meV/f.u. (relative to monoclinic). The considerably different values for the energy differences result from different density functionals where the local density approximation (LDA) typically gives smaller and the generalized gradient approximation (GGA) larger energy differences. Further theoretical uncertainty arises from d-orbitals in standard DFT which can be better described with more sophisticated methods like DFT+U or HSE (some examples are shown in Fig. 1a). Results obtained with this method are consistent and lead to conclusion that the ferroelectric phase has to be stabilized relative to the monoclinic phase by certain mechanisms. Important for phase stability is the expression of the free energy

$$F = U - TS, \quad (1)$$

where $U_0$ in $U=U_0+U_{zero}$ is the total energy, T the temperature, $U_{zero}$ is from zero point vibrations and S the vibrational entropy. Dependence on electric field, stress and surface energy will be discussed in section 2.1. Figure 1a shows that the vibrational entropy lowers especially the free energy of the tetragonal phase with temperature, which should therefore prevail under annealing conditions. For zirconia, the energy difference between the competing ferroelectric and tetragonal phase becomes very small and the free energy (Fig 1b) nearly vanishes. In the solid solution $Hf_{1-x}Zr_xO_2$ the free energy differences can be linearly interpolated between $HfO_2$ and $ZrO_2$. Thus, $ZrO_2$ can be considered as a dopant stabilizing the ferroelectric phase, although with a competing tetragonal phase.

In summary, this result describing the ferroelectric phase as a metastable phase matches well with the fact that it only exists under certain circumstances as there are film thickness and grain size, dopants, impurities and as speculated stress and or electric fields. In pure hafnia, the ferroelectric phase free energy is about 50 meV/f.u. relative to the monoclinic ground state such that it can be only stabilized in extremely thin films [27]. On the other hand, in pure zirconia the tetragonal phase is only a few meV/f.u. close to the ferroelectric phase such that an external electric field is sufficient to induce its stability [28, 29]. Around the 50% mixture of hafnia and zirzonia the ferroelectric phase is very stable under thin film conditions (up to a thickness of about 25 nm).

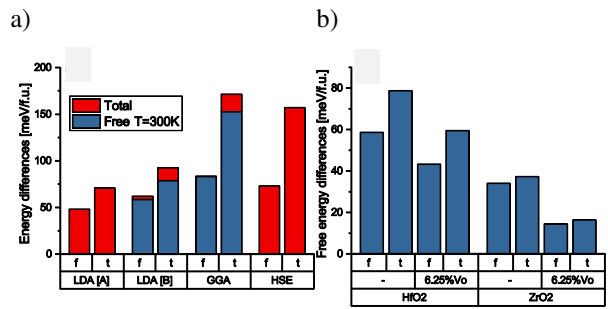

*Figure 1.* (a) Total and free energy differences of $HfO_2$ ferroelectric (f) and tetragonal (t) phase relative to monoclinic for different DFT methods, (b) free energy of $ZrO_2$ and $HfO_2$ as well as the effect of 6% oxygen vacancies ($V_O$) on free energy. LDA [A] values are from Ref. [30], LDA [B] from [29], GGA from [31] and HSE from [32]. Values for b) are calculated based on Ref. [29].

$Hf_{1-x}Zr_xO_2$ is a model system for ferroelectric phase stabilization with a dopant. Namely, for a thin film (of about 10 nm), increase of the concentration of an appropriate dopant for a few anionic % lowers the free energy of the ferroelectric phase as well as of the tetragonal and cubic phases relative to the monoclinic one. For the four valent Si there is a concentration range from 3-5cat% (cationic ratio Si/(Hf+Si)) favoring the stabilization of the ferroelectric phase. For the three valent dopants N, Al, Y, Gd and La and for the two valent Sr and Ba [32], concentration ranges from few to more than 10 cat % have been found. Beyond this range, the film becomes tetragonal for the four valent and cubic for the three and two valent dopants. Even though this topic is well suited for DFT simulations, apart from Fischer et al. [33] and Lee et al. [34] (where dopants stabilizing the tetragonal high-k phase were searched), no systematic study of phase stabilization with dopant has been reported so far. There are several reasons for the slow progress: whereas tetragonal or cubic phase stabilization clearly dominates beyond some doping range, ferroelectric phase stabilization seems to be possible only in a distinct concentration window requiring more accurate calculations. The introduction of various dopants might be accompanied by the formation of oxygen vacancies providing the charge compensation, and consequently requiring the calculation of a large number of structures to find the most favorable configuration. Finally, the resulting total energy or more demanding free energy calculation results do not fully explain the phase stabilization in contrast to the experiments. Obviously, either the DFT calculations are wrong, or some further stabilization mechanism in addition to the dopant is involved. We believe that surface or interface energy in polycrystals, discussed in section 2.2., is a major effect closing the gap between simulation and experiment. Such a mechanism was already proposed in [33] to explain the stabilization of the tetragonal phase in thin films by DFT data correctly.

Charge compensation in doped, ferroelectric hafnia required to avoid excessive leakage currents and charge build-up, might work in pairing dopants with oxygen vacancies, but still needs to be demonstrated in detail. By calculating the occupation states of the defect levels in the band gap, the efficacy was shown for monoclinic hafnia and Y [35], Al [36], Gd [37] or Ba [38] as a dopant.

Oxygen vacancies, $V_O$, in hafnia without compensating dopants are causing leakage current and charge build-up. The defect structure of the monoclinic [39] phase has been researched intensively and reveals the double charged $V_O^{++}$ with unoccupied states about 1eV below the conduction band as the most stable structure for zero applied field.

Not only the paired, but also the unpaired vacancies might contribute to the desired phase stabilization. Here, some preliminary results have been published [40], shown in Fig 1b. 6.25 ani.-% $V_O$ in $HfO_2$ lowers the ferroelectric phase by 15 meV/f.u. and the tetragonal phase by 20 meV/f.u. relative to monoclinic phase. Interestingly introduction of $V_O$ favors the stabilization of the tetragonal phase with respect to the ferroelectric. This leads, leading to the typical concentration window with the tetragonal phase dominating at large concentrations. If oxygen vacancies play a role in the phase transformation observed during field cycling, however, they have to be present somewhere with a concentration of a few percent (not all $V_O$-s are active from the electrical point of view).

**2.1 Phase transitions within binary oxides**

A phase transition of ferroelectric hafnia into the tetragonal phase is desired when pyroelectric or electrocaloric effects are being exploited, controlled by temperature T and electric field E in a thermoelectric cycle. In contrast to the pyroelectric and electrocaloric applications, phase transitions are highly undesired in memory application. Nonetheless, for reaching the desired phase stability it is important to investigate its dependencies. The free energy for each phase $\phi$ can be calculated as

$$F_\phi = U_\phi - TS_\phi + \Omega_{0,\phi} D_\phi E_\phi + \Omega_{0,\phi} \sigma_{ij,\phi} s_{ji,\phi} + \Gamma_\phi, \quad (2)$$

where $\Omega_{0,\phi}$ denotes the volume of a formula unit of the phase, $E_\phi$ the electric field, D the dielectric displacement, S the vibrational entropy, $\sigma_{ij}$ and $s_{ij}$ stress and strain tensor, respectively and $\Gamma_\phi$ a surface energy contribution.

The calculated temperature dependence from $U_{\phi,zero}$ - $TS_\phi$ with the electric field Ez in the polarization direction is shown in Fig 2. It can be seen that a 150 K decrease in temperature results in about a 4meV/f.u. increase of the tetragonal energy difference. This allows the ferroelectric phase stabilization at the phase boundary to the tetragonal phase by lowering the temperature, as shown in [41].

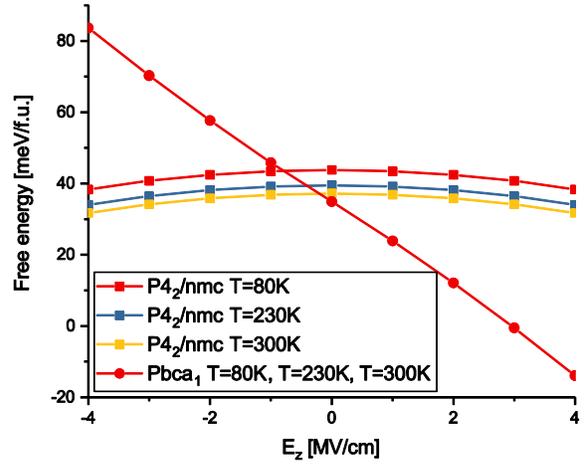

*Figure 2.* Free energy as a function of the electric field in the polarization direction $E_z$ and the temperature for the tetragonal and ferroelectric phase of $ZrO_2$ related to the monoclinic phase. The calculated values have been taken form [29].

The impact of an electric field on the free energy was studied by Materlik et al. [29] to explain the antiferroelectric (AFE) behavior in thin $ZrO_2$ as an electric field induced phase transformation and by Batra et al. [42] to explore phase stabilization by internal fields. For a field strength of 1 MV/cm the value of $\Omega_0 DE$ is 10meV/f.u., Fig. 2 shows the calculated scenario of AFE behavior as field induced phase transition which fits to the observation of [41] in thin zirconia. These findings are of high importance for the stabilization of the distinct material into novel AFE non-volatile memories [22,23].

From Eqs. (1 and 2) one would expect a temperature induced phase transition to the tetragonal or cubic phase at a fixed temperature (the Curie temperature). In Si doped $HfO_2$ [40] a large interval of transition temperatures has been found (see Figure 3). This has been explained with the surface or interface energy contribution $\Gamma_\phi$ which differs for grains of different size in the film. Since the derivative of $P_r$ with respect to T is the pyroelectric coefficient (sum of proper and giant, phase transition induced contribution), the perspective is to tailor the temperature range and magnitude of the pyroelectric coefficient with Si-doping as well as the grain size distribution [40].

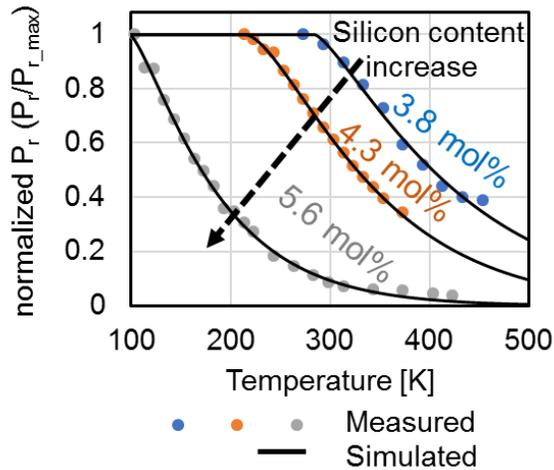

*Figure 3.* Remanent polarization in Si:HfO$_2$ for different doping cat% as a function of temperature.

A last constituent of Eq. (2) to be discussed is the energy of stress or strain. Stress has been claimed as the major effect to stabilize the ferroelectric phase [30]. The required stress depends on the energy difference value to be overcome (see Figure 1.), which fluctuate for different DFT methods. In [30] it was shown that a planar, compressive strain of 1% (which corresponds to a stress of about 3 GPa) is required for stabilization of the ferroelectric phase. Materlik et al. [29] calculated the required stress to be 8 GPa which seems unrealistic high and therefore, a surface energy mechanism was proposed. Finally, Batra et al. [42] assumed a combination of 2 % compressive strain together with an internal electric field of 1.5 MV/cm. Although stress or strain can play a role in stabilization of the ferroelectric phase in hafnia and zirconia, it seems that the first order effect must lay in some other phenomenon.

So far, we have treated phase stability in the thermodynamic picture with the free energy. If large barriers are involved in a transition, kinetic effects may become visible. The transition barrier from the tetragonal to the ferroelectric phase has been calculated by several authors [31][30][29] being about 30 meV/f.u. and of first order. The transition barrier from monoclinic to ferroelectric has not been reported, but should be much larger around 210 meV/f.u., according to [43]. This raises the question whether a field-induced transformation from monoclinic to ferroelectric is possible, however there is no clear evidence so far.

Polarization reversal is a special case of phase transition. Huan et.al. [31] have proposed that the tetragonal phase is at the intermediate state between both polarities of the ferroelectric phase (Fig. 4).

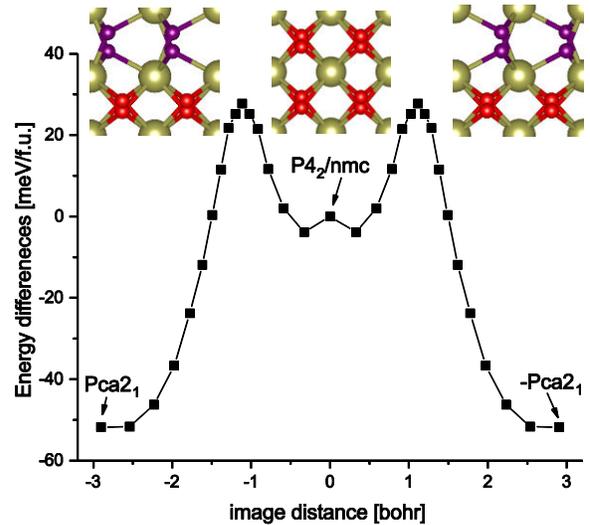

*Figure 4.* DFT calculation of the transition path from the ferroelectric phase via the tetragonal to the negative ferroelectric phase. Energy differences are produced with pseudopotentials using the String method. The image distance is calculated as the difference of the Frobenius norm of all coordinates.

A simple first order estimate for the coercive field, taking the smaller polarization at the transition state into account [30] results in a value of 1.2 MV/cm which is in good agreement with the experiments. A more sophisticated approach calculating the barrier height under the applied field has not been reported so far.

All the calculations for kinetic effects reported for Hafnia have not considered long range field effects which are present in perovskite ferroelectrics for which effective Hamiltonian models have been developed [44].

**2.2 Stabilization of the Ferroelectric Phase: Surface or interface energy model**

The thickness dependence investigation of the free energy shows that the decrease of the hafnia film thickness favors the ferroelectric phase, whereas the underlying phase in thick films becomes unstable. In contrast to the typical observation in perovskite based ferroelectrics this trend of binary oxides is very favorable for the integration of ferroelectrics in ultimately scaled devices. It should be noted that there are probably exceptions for the rule that can be concluded from the experimental results shown in Fig. 5. Besides the dopant concentration window of stability there is a maximum thickness for which ferroelectric phase in $Hf_{0.5}Zr_{0.5}O_2$ is stable, with an upper boundary of about 25 nm. Even though there is an indication of decrease of $P_r$ at film thicknesses of 5.5 nm, however thinner films have not been investigated so far.

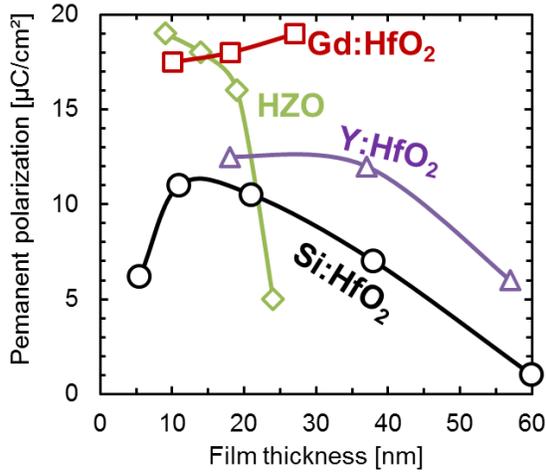

*Figure 5.* Thickness dependence of $P_r$ for different doped hafnia.

A surface or rather interface energy related size effect defining the crystallographic phase of individual grains has been proposed by Materlik et al. [29] and Künneth et al. [45], respectively assuming cylindrical grains of radius r and height h = t (where t denotes the film thickness). The free energy contribution to a grain is

$$\Gamma_\phi = \frac{\text{surface area}}{\text{volume}} \gamma_\phi = \left(\frac{2}{t} + \frac{2}{r}\right)\gamma_\phi \qquad (3)$$

being a function of d as well as r. Hence, free energy contribution to a grain in average is a function of the grain radius distribution f(r,t,x), where x is the dopant concentration.

Crucial in the model are the values of surface or interface specific energies $\gamma_\phi$. The suppression of the monoclinic phase with a size effect requires $\gamma_m > \gamma_t$ and $\gamma_m > \gamma_f$ (where $\gamma_m$, $\gamma_t$, $\gamma_f$ denote the interface or surface specific energy of monoclinic, tetragonal and ferroelectric phase, respectively). The former condition has been experimentally confirmed and the condition $\gamma_t \approx \gamma_f$ is reasonable considering the similarity of the phases. Furthermore, the term $\gamma_\phi(x)$ depends on the dopant concentration x. In [29] a free energy model for $Hf_{1-x}Zr_xO_2$ was developed with a linear dependence of the crystal free energies and $\Gamma_\phi$ on x (due to the good solubility of $HfO_2$ in $ZrO_2$ and vice versa). Hence, resulting in x linear free energies $F_\phi(x)$ for each phase where the intersections define the phase boundaries. Depending on the hafnia-zirconia mixtures, naturally a window of stable ferroelectric phase in x appears, varying with thickness, as shown in Fig. 6. This model is considered to describe the thickness dependence of the ferroelectric phase in $Hf_{1-x}Zr_xO_2$. Generally, for other doped hafnia films, a part of the stabilization is achieved with the dopant and the thickness dependence could be modeled with the surface energy contribution. In [45] the model reported in [29] has been refined with proper inclusion of the grain size distribution. By using total energies from DFT, data can be reproduced quite well, however, the values for $\gamma_\phi(x)$ are taken from a fit. Batra et al. [42] have used DFT calculations for $\gamma_\phi$ (Hf) and obtain a size effect, although no good fit to data, which is no surprise due to the present DFT error present due to the chosen DFT functional (see Figure 1 for details).

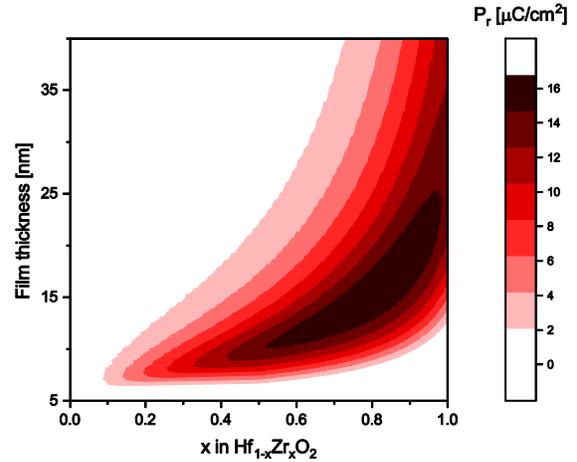

*Figure 6.* Remanent polarization in $Hf_{1-x}Zr_xO_2$ as function of concentration x and film thickness t modeled with a surface energy model reported in [45].

## 3 Physical Mechanisms Model behind field cycling of ferroelectric memory elements

After the detailed discussion of the factors determining the origin of the ferroelectricity in hafnia based thin films, in the following a central reliability issues of ferroelectric-based memories will be reviewed. Even though the retention of presented memory architectures can be extrapolated to a 10-year specification target, both hafnia based FeRAM and FeFET memory architectures suffer of rather limited endurance. Endurance characteristics of the FeFET and FeCap devices are shown in Figure 7a-b. In contrast to the FeCap case, where typically hard breakdown occurs before the closure of the memory window (MW), continuous PRG/ERS stress of the FeFET eventually results in complete closure of the MW due to the charge trapping. The transfer characteristics of transistor characteristics remain measurable [46], which means that the device still works as a transistor, but its memory properties are lost.

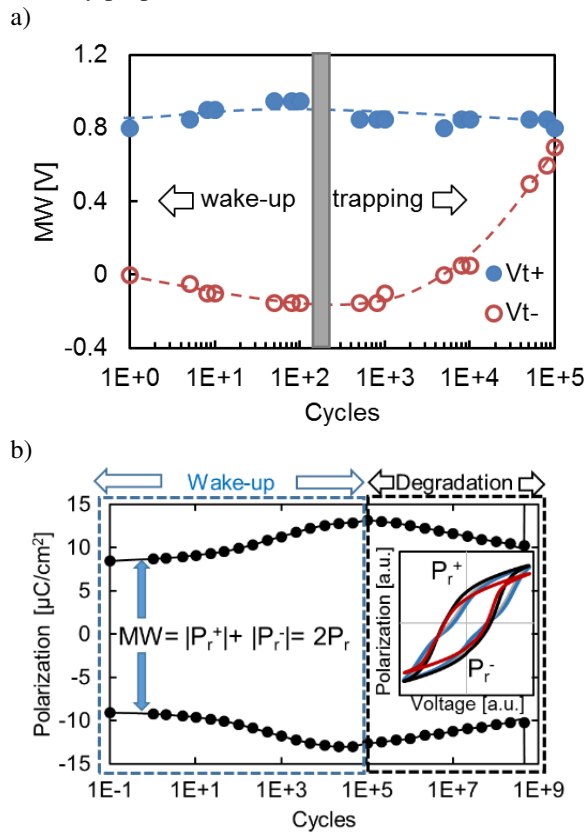

*Figure 7.* Typical endurance characteristics of the a) FeFET measured on the multi-structure with program erase condition of -5/4.5 V. b) FeCap recorded at 10 kHz read frequency and at 100 kHz cycling at 3.3V.

To address the question of the root cause of memory window change and degradation, a current understanding of field cycling evolution of these two polarization based memories will be reviewed. Moreover, endurance characteristics of the ferroelectric capacitor can be splitted into two distinct lifetime stages (see Figures 7a and b):

a) "Wake-up": The increase of the remnant polarization $P_r$ corresponds to an opening or de-pinching of the pristine pinched hysteresis loop with field cycling [47][48];

b) "Fatigue": After a certain number of cycles, the remanent polarization starts decreasing, which results in a closure of the memory window due to fatigue mechanisms [49].

The underlying physical mechanisms behind wake-up and fatigue of conventional ferroelectric materials like PZT and SBT were studied in detail during the past 50 years. A large number of publications proposed numerous scenarios dealing with increase (wake-up) and decrease (fatigue) of the polarization in perovskite-based materials. Different studies of wake-up on the PZT material system were performed and reported previously [47][50][51]. Lou et al. in [49] presented very detailed and systematic reviews of possible fatigue mechanisms. Even though these studies suggested that a combination of several mechanisms (domain de-/pinning [52], seed inhibition [53], and formation of a passive/dead layer [54]) could be responsible for the device wake-up and fatigue neither one gave a detailed modeling approach of the underlying effects. The first comprehensive modeling study of the mechanisms responsible for wake-up and fatigue was reported in 2016 on the hafnia based ferroelectric memories [55].

Before the actual modeling is described, a brief review of the phenomena occurring due to field cycling is given. These are extracted from hafnia based ferroelectric capacitors utilizing various structural and electrical characterization techniques. The pristine stage of the lifetime of ferroelectric capacitors is typically characterized by a double current peak feature in the dynamically measured transient current [55,56]. Since the polarization is the integral of the transient current, the double current peak characteristics is transformed into a pinched hysteresis loop in the polarization-voltage domain [57]. The continuous stress (field cycling) of the device results in a merging of the double peak characteristics and corresponding opening of the initially, pinched hysteresis loop [56]. During that stage, it was reported that the leakage current stays rather constant [55] whereas the transmission electron microscopy investigations pointed on phase transitions within the material [55, 58]. Further cycling of the FE capacitor results in a broadening of the current peak and as a consequence decrease of the $P_r$ amplitude (fatigue stage) and is followed by the increase of the leakage current magnitude [55, 59].

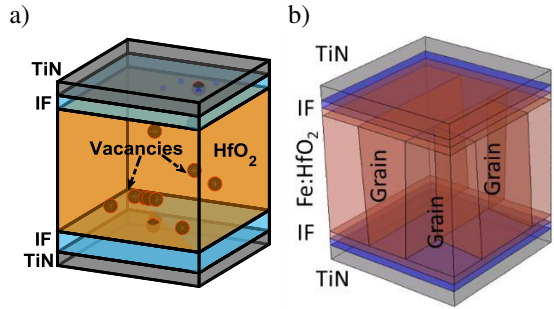

*Figure 8.* 3D geometry of the a) FeCap structure implemented in MDLab software used for modeling of the vacancy diffusion process and b) FeCap multidomain model developed in TCAD.

In the following a review of the structural and physical properties that have to be accounted for in the model is given. First the impact of the two additional interfacial regions on the electrical behavior of the MIM (TiN/X:HfO$_2$/TiN) capacitor (where X is the dopant) has to be considered. In addition to the the TiO$_x$N$_y$ and TiO$_x$ interfacial layer formation due to oxygen scavenging from the HfO$_2$ and nitrogen diffusion, results of the TEM study [55, 58] revealed parasitic tetragonal regions of the HfO$_2$ film towards the electrodes. Due to their tetragonal nature, these regions consist of non-switching (from the ferroelectric point of view) transitional material (T-HfO$_2$) [55, 60, 61]. Therefore, the complete device stack consists of TiO$_x$N$_y$/TM-HfO$_x$/FE-HfO$_2$/TM-HfO$_x$/TiO$_x$ sandwiched between two TiN electrodes (see Fig. 8 for details). Further this material intermixing that results in formation of a thin ~1 nm thick sub-stoichiometric TiO$_x$ and TiO$_x$N$_y$ regions may affect the effective k-value of the stack. The interfacial regions are characterized by a large defect density (oxygen vacancies; V$_o$). According to the TEM study, these sub-stochiometric low-k TiO$_x$ regions are accompanied by a roughly 1.5±1 nm thick HfO$_2$ interface characterized with a higher k-value due to the presence of a tetragonal phase. These two factors significantly impact the voltage distribution inside the dielectric layer. Hence, a superposition of the lower dielectric constant and the electron trapping at interfacial defects can results in either a local increase or decrease of the effective field, which influences the resulting polarization of the ferroelectric film.

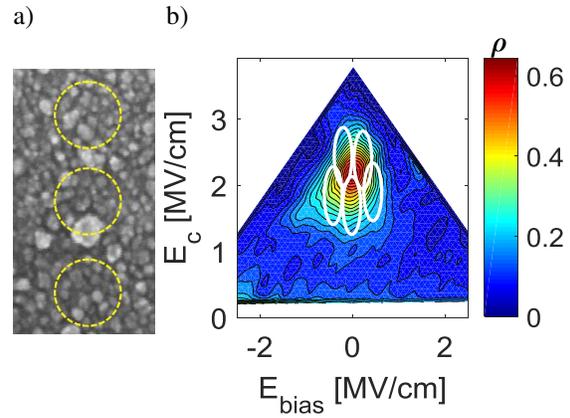

*Figure 9.* a) SEM top-view of a polycrystalline hafnia film: Granular structure of the capacitor device where each FeCap (sketched circled area) has a random grain distribution. b) Corresponding domain separation depicted on a Preisach density plot obtained using a FORC technique.

In the next step, the polycrystallinity of the HfO$_2$ layer has to be explicitly taken into account. In order to stabilize the desired crystal phase and therefore the ferroelectric properties, hafnia-based devices are crystallized using an 800-1000 °C anneal for 20 seconds in nitrogen atmosphere which induces a granular, polycrystalline morphology of the film. Scanning electron microscopy micrographs (see Fig. 9a) measured on the bare HfO$_2$ oxide film after top electrode removal showed that a specific device comprises a random distribution of grain sizes. Moreover, the total area of the investigated capacitor (33000 μm$^2$) consists of hundreds of thousands of grains which all might have a slightly different domain orientation, coercive field or remnant polarization. To address the polycrystalline nature of the film, the modeled hafnia ferroelectric layer is divided into randomly distributed grains and two interface layers next to the electrodes. The granular morphology and grain boundaries play very important roles since they behave as the preferable locations for the accumulation of the V$_o$-s. In order to model the variability in the resulting switching characteristics, each of the modelled grains should represents an averaged ensemble of domains with similar properties. Each ensemble is defined by a distinct coercive field, remanent polarization value and k-value. The coercive fields used for the ensembles can be chosen based on the switching densities observed in the First Order Reversal Curve (FORC) [62] distributions (see Fig. 9b). In ref. [63, 64] it was reported that grains can be correlated with discrete ferroelectric domains and switching events within highly scaled FeFETs (details and modeling approach will be discussed in section 4.). Further to address the phase transition with field cycling, different phases of the doped hafnia, which can coexist inside the same grain: a) the orthorhombic phase, responsible for the ferroelectric switching; b) monoclinic, which are not active from the ferroelectric point of view and c) tetragonal phases, responsible for AFE behavior should be taken into account [41, 55].

To account for the behavior of a ferroelectric, the model has to take the history dependent charge-voltage relationship of the ferroelectric [55, 65, 66] into account. Ferroelectric behavior can typically be simulated using either a Preisach model of hysteresis or the Landau-Khalatnikov theory of phase transitions. The differences between those two approaches and the need for the L-K approach in case of ultimately scaled devices will be discussed later in sections 4.2 and 4.3. The hysteretic behavior of the ferroelectric is most commonly represented using a Preisach-based, numerically stable model of the hysteresis [67] which is readily available within the commercial TCAD packages used for device simulations [66] and is governed by:

$$P = c \cdot P_s \cdot tanh\left[\frac{1}{2E_c} ln\frac{P_s+P_r}{P_s-P_r}(E \pm E_c)\right] + P_{off}, \quad (4)$$

where $c$ represents the proportionality constant, $P_s$ the saturation polarization, $P_r$ the remnant polarization, $P_{off}$ the offset polarization, $E$ the electric field and $E_c$ the coercive field.

(details of the implementation and model derivation can be found in refs [55, 65, 66, 67]).

In the next step, to account for a parasitic influence of the dielectric leakage on the ferroelectric switching a Nasyrov-type multiphonon mediated trap-assisted-tunneling (TAT) model was implemented [68, 55, 66]. Every high-k material contains a significant number of defects such as dislocations and impurities which act as electron traps. To take those trap states into account they are modeled as single level acceptor and donor traps within the TCAD software.

These oxygen vacancies actively participate in the TAT current and are responsible for the local field modifications due to charge trapping.

Further, to access the possible material diffusion due to the application of high electric fields, the diffusion and recombination of oxygen ions and vacancies have to be calculated through a kinetic Monte Carlo (kMC) model implemented in the MDLab package [69, 70]. Recombination rate $R_D$ and diffusion rate $R_R$ can be calculated using the following equations:

$$R_D = \nu \cdot exp\left(-\frac{E_{A,D}-Q\frac{\lambda}{2}E_{eff}}{k_BT}\right); \quad (5\text{-}1)$$

$$R_R = \nu \cdot exp\left(-\frac{E_{A,R}}{k_BT}\right), \quad (5\text{-}2)$$

where Q is the charge of the diffusing space, λ is the jump distance, $E_{A,D}$ is the activation energy for ion/vacancy diffusion, and $E_{eff}$ the effective field along the jump direction, $E_{A,D}$ the activation energy (energy barrier) given by $E_{A,D}$=0.7 eV and $E_{A,D}$=1.5 eV for $O_2$-ions and positive $V_o$-s ([71]), respectively. $E_{A,R}$ represents the activation energy for recombination between complementary species.

In the same study [55, 59 72], recombination and diffusion processes were coupled to the charge transport model.

At the end, to emulate the stress-induced defect generation, the thermochemical bond breakage model implemented in the MDLab software [69,72, 73] has to be considered. Hafnium oxygen (Hf-O) bond breakage generates an oxygen vacancy, which is an electrically active defect (contributes to the TAT transport), and an oxygen ion. The vacancy generation rate G is described by equation (6).

$$G = \nu \cdot exp\left(-\frac{E_{A,G}-bE}{k_BT}\right), \quad (6)$$

where ν is the bond vibration frequency, b the bond polarization factor, E the electric field, $E_{A,G}$ the activation energy for generation of an ion-vacancy pair, $k_B$ the Boltzmann' constant and T the temperature.

### 3.1 Internal bias field and Wake-up

In ref. [62] it was reported that pristine double current peak behavior could originate from different internal bias fields within the device stack. In the same study it was assumed that different local internal biases or a different effective field manifestations on certain grains and even portions of the device may occur due to a non-uniform defect distribution, polycrystallinity of the annealed film and consequent k-value non-uniformity [55, 65]. As reported in the previous section, the origin of the observed wake-up in the ferroelectric memory is attributed to the progressive decrease of the local internal bias fields. Pesic et al. reported that the root cause for this this decrease is an $V_o$ defect redistribution that can change their charge state, and/or induce partial phase transitions and k-value changes within the layer [55]. Within this study defect evolution with cycling was studied by monitoring the static leakage current in different lifetime stages of the device. It is very important to note that the density of defects evaluated from the leakage current corresponds mainly to those located at the grain boundaries, where they account for the main contribution to the leakage current [69]. More precisely, the leakage current through the grain boundaries is more than tenfold higher than through the grain itself. Thus, the diffusion of the vacancies though the grain would keep the leakage current constant during the wake-up stage. The role of O vacancy/ion diffusion is also consistent with the fact that $HfO_2$ based devices are known for their high oxygen mobility [57]. This claim is strengthened by the fact that the electric field applied to $HfO_2$-based ferroelectrics is at least one order of magnitude higher than the fields typically applied to PZT based films [57]. Furthermore, a recent transmission electron microscopy study confirmed the generation and movement of oxygen ions and vacancies within 10 nm thick $HfO_2$ film under similar operating conditions [74]. Starschich et al. proved resistive and ferroelectric switching operation in the same $Y:HfO_2$-based device [75]. It is widely accepted that oxygen vacancy motion is required for resistive switching in $HfO_2$ [70].

To address the influence of cycling and dynamics within the stack as well as its influence on the field distribution (predominantly on internal bias fields), the diffusion and recombination of oxygen ions and vacancies through a kinetic Monte Carlo model implemented in the MDLab package [70, 76] was simulated. Diffusion rates (equation (5-1) were calculated by considering activation energies for ion and vacancy diffusion of 0.7 eV and 1.1 eV respectively [74, 77]. Based on the previous discussion, the constellation of the pristine state within the model was set so that the transitional regions at the electrode interface have much higher defect density. Moreover, based on a TEM study [55] the pristine condition (and internal bias) was generated considering the non-switching interfaces which possess different k-value with respect to the switching fractions of the film which consequently affects the distribution of the local fields. In addition to the low-k value this additional local internal bias field component was created by highest density of the vacancies in the interfacial regions [55, 65].

This scenario was used to simulate the preferable movement of the O vacancies driven by the applied electric field. Simulation of bipolar stress cycling revealed that O vacancies redistribute uniformly within the grain [55]. In addition to this drift/diffusion process, recombination can occur. Due to the recombination of vacancies and interstitial ions the internal bias field decreases. As a result, uniform field distribution is created within the stack, resulting in homogeneous switching of all domains within the device [55]. Temperature accelerated wake-up experiments reported in [55] motivated a simulation [55] which consistently pointed out that increase of temperature, results in facilitated vacancy diffusion. Hence, yielding more mobile vacancies and faster achievement of a uniform electric field within the ferroelectric [55].

In the same study, it was reported that field driven vacancy distribution during the wake-up stage increase the leakage current component through the grains, however this is insignificant compared to the total current governed by leakage thought the grain boundaries.

The interfacial regions are considered non-switching in the pristine state. Hence, they act as a passive (dead) layer [55]. The monoclinic (non-switching) phase was modeled by setting the grains to a low-k (with respect to the orthorhombic grains), dielectric state.

a)                                b)

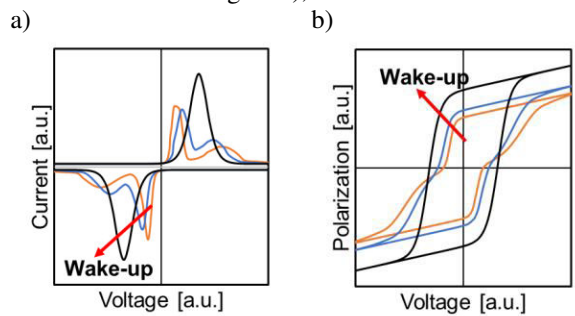

*Figure 10.* Simulated wake-up of the ferroelectric MIM capacitor comprising pristine, intermediate and woken-up a) current-voltage and b) polarization-voltage characteristics.

Using the Preisach model [67], a double peak current-voltage characteristic as well as a pinched hysteresis loop was successfully simulated (see Figure 10). In-addition domain de-pinning was represented by removing the charges previously located at the electrode interfaces, whereas the phase transformation was included by changing of the respective k-values and setting the portions of interface as well as the previously non-switching grains to a ferroelectric active state. These modifications resulted in merging of the current peaks or in other words the opening of the hysteresis curve (Figure 10), that completely emulates the behavior of the real device.

It should be noted that even the film is woken-up, due to the polycrystalline nature of the film, a slight deviation of experimentally determined $P_r$ from the theoretical, calculated remanent polarization were usually reported. On the other side the remanent polarization calculated for hafnia and zirconia from various authors and from [78] assuming different unit cell sizes from doping with values ranging around $P_{calc}$=50 μC/cm$^2$ and a single polarization direction. Assuming isotropic orientation of grain in a polycrystalline film, the expected average remanent polarization is

$$\overline{P_r} = \frac{\pi}{8} P_{calc} \approx 21 \frac{\mu C}{cm^2} \quad (7)$$

Which matches experimental values very well.

## 3.2 Fatigue

The field cycling endurance of a ferroelectric capacitor may be limited by two aspects: a) degradation of the dielectric which eventually leads to a hard breakdown and b) reduction of the switched polarization leading to a closure of the ferroelectric memory window. The later aspect is typically referred to as fatigue.

First, we want to focus on the time dependent dielectric breakdown (TDDB). Detailed comparisons of the DC and AC TDDB lifetime of the ferroelectric capacitors was reported by Masuduzzaman et al. [79, 80]. In these reliability studies, it was shown that the traditional breakdown theories for gate oxides are not sufficient to explain the AC TDDB stress dependence of ferroelectric materials. Devices under test comprised a 70-nm thin ferroelectric PZT film. Similar to the standard (non-ferroelectric) oxides DC TDDB lifetime ($T_{BD}$) followed a power law with increasing voltage, characterized with a rather small Weibull slope ($β_{DC}$~1.5). Moreover, employing the conductive PFM technique it was shown that the breakdown (BD) spot was always located within the grain boundaries. However, when transferring to AC stress, authors observed drastic decrease of the $T_{BD}$ followed with significant change of the Weibull slopes. The Weibull slope $β_{AC}$ for AC stress was significantly increased~9.1. The reason for this behavior can be explained by modeling of hot ion (called hot atom by the authors of [79,90]) degradation which indicated that the increased Weibull slope originates from the breakdown occurring through the grains in contrast to the previously discussed DC induced breakdown taking place at the grain boundaries.

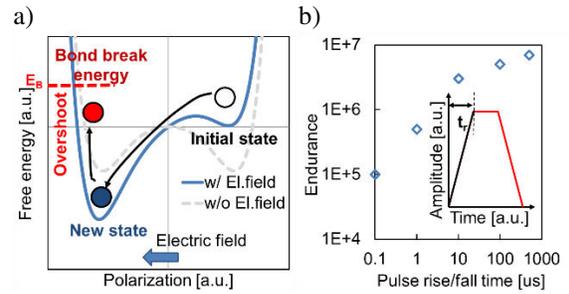

*Figure 11.* Hot ion degradation concept. a) Double-well energy potential for each switching event, b) Breakdown-endurance dependence on the rise/fall time of the stress pulse. (inset) Pulse with variable rising edge.

The same authors reported a hot ion overshoot to be responsible for the degradation and bond breakage due to alternating voltage stress and subsequent ion crossing of the barrier within the double-well potential (known from thermodynamic Landau theory) describing ferroelectric materials. Equation (6) describing bond breakage can be modified including the local polarization $p$ in such way that the bond breakage energy is represented as

$$E_{A,G} = E_{A,G} - pE, \quad (8)$$

Since polarization is a dynamic parameter it is described with the following differential equation

$$A \frac{d^2}{dt^2}(p) + B \frac{d}{dt}(p) - F(p) = E(t) \quad (9)$$

where A and B are inertial and damping coefficients, respectively and $F(p)$ denotes internal field originating from double well ferroelectric potential and corresponding Landau coefficients (α, β, γ) [79, 80] and is described with equation

$$F(p) = -\frac{d}{dt}\left(\frac{\alpha}{2}p^2 + \frac{\beta}{4}p^4 + \gamma \nabla p\right) \quad (10)$$

Combining equations (6) to (10) it was shown that an overshoot of the ion can be caused if pulses with ultra-steep edges are applied. During the change of the polarization state, the ion has a kinetic energy high enough, not just to overcome the barrier but also to climb to the opposite wall of the energy potential eventually reaching the bond breakage energy (due to this overshoot). This bond breakage results in the generation of the defects, i.e. $V_o$-s and interstitial O ions in this case, which finally lead to dielectric breakdown. Simulations by Masuduzzaman and coworkers [79, 80] show that with the increase of the length for the rising/falling edge a strong decrease of the overshoot and corresponding increase of the lifetime of the device was obtained. In order to examine the applicability of the concept, endurance experiments were performed on the $HfO_2$ based MIM capacitors. The plateau width of the pulses was kept constant (1 µs) whereas the rising/falling time of the stress pulses were varied from 100 ns to 1 ms (Figure 11b, inset). Indeed, the increase of the rise/fall time of the stress pulses resulted in improved endurances. Even though the experiment was successfully repeated, a detailed statistical study is needed in order to draw a final conclusion about the applicability of the method.

Concerning the switching induced degradation, in study by Pešić et al. [59] endurance of $HfO_2$ based ferroelectric MIM capacitor devices recorded while stressing with unipolar pulses and compared to endurance monitored while stressing with bipolar pulses. Similar to the results of the breakdown study by Masuduzzaman et al. [79, 80] it was reported that only the alternating switching, i.e. a continuous change of the polarization state, results in a degradation and consequent reduction of the MW. In contrast the unipolar stress does not influence the MW significantly [59, 81, 82]. Leakage current defect spectroscopy proved that independent of the polarity of the unipolar stress pulses, both leakage current and memory window stayed constant. Analogues to the hot ion degradation model, it was concluded that alternating polarization switching itself (continuous ion displacement within the ferroelectric crystal lattice) induces the endurance degradation and fatigue. A direct correlation between the increase of the defect concentration extracted from the charge transport model of the ferroelectric capacitor and the degradation of the MW during the fatigue stage was reported [59]. Beside the degradation of the bulk of the device it was reported that faster O vacancy defect generation occurs within the $TiO_x$ interface next to the electrodes [59]. Since the bond breakage heavily depends on the applied field and the activation energy, usually given by the stoichiometry of the film, these trends can be concluded from equation (6) describing the thermochemical bond breakage.

These sub-stoichiometric, non-switching interface regions characterized by a high number of defects and a lower permittivity resulting in higher field drop, yield an increased factor of degradation [65] and result in faster bond breakage. The subsequent charge trapping on this generated defects significantly influences the local field distribution across the device stack. The field over the interface increases, while it reduces inside the ferroelectric bulk [59, 65]. As a consequence, the ferroelectric layer (switching/active part of the device) experiences a lower field, which leads to a reduced number of switching domains and decrease of the MW.

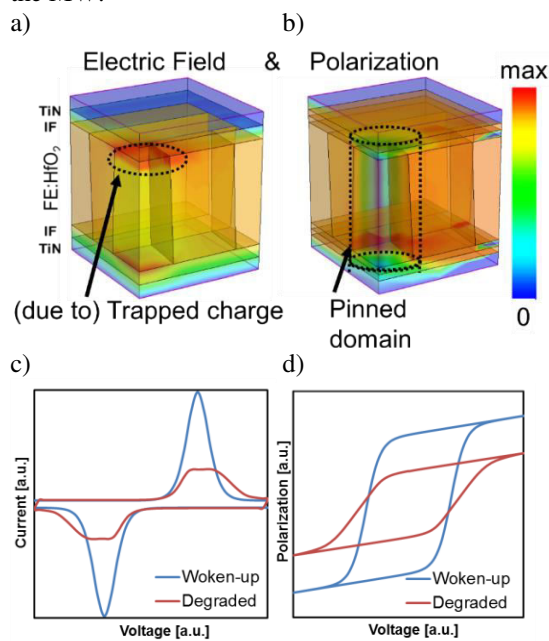

*Figure 12.* Charge trapping influence on the electric field: Polarization response and corresponding domain pinning due to the charge trapping of a) woken-up and b) fatigued stack. For the reasons of clarity single domain pinning is illustrated. Interface, IF comprises parasitically grown $TiO_x$ as well as non-switching portion of $HfO_2$. Comparison of the woken up and degraded state obtained by simulation: c) Current and voltage and corresponding d) Polarization and voltage response.

In refs. [55, 59] the modeling of the degradation and its influence on the ferroelectric switching was performed. Extracted defect concentration and obtained defect distribution were used as input for the 3D grain boundary model of the MIM capacitor reported in [55] and discussed in section 3.1. The degrading influence on the field distribution across the stack and the resulting change of the current-voltage and polarization-voltage characteristic is depicted in Figure 12a-b. Pure electrostatic influence of shielding (charge trapping increases field over the interface and decreases over the ferroelectric), is accompanied with the domain pining. The trapped charge creates dipoles impeding the switching of the domains, which results in a partial or even complete pinning of the domains [59]. For simplicity, the simulated example given with Figure 12 shows charge trapping within one corner of the top interface region, which alters the local field distribution and pins the domain in the current grain (bounded with the dashed cylinder). As in the measurements [55], also the simulated current-voltage traces resulted in broadened peaks with decreasing magnitude. As consequence of the altered field, the simulated polarization hysteresis was characterized by smoother transitions and lower $P_r$ (see Figure 12c-d).

## 4 Towards ultimately scaled 1T-Memory Architectures

In this section an overview of the current progress in modeling approaches utilized in investigation of ultimately scaled hafnia based 1T architectures is presented. Even though various studies reported modeling of the memory window of FeFETs, rather few addressed the limitations caused by switching kinetics and nucleation of the domains in ultimately scaled devices. Exactly these properties together with multigrain operation and multi-level-cell (MLC) possibilities will be addressed. Moreover, an overview of the progress in understanding on the neighboring cell disturb through the mix-mode simulations will be reviewed. For this purpose, a 2D multi grain model of the FeFET device was utilized (see Figure 13.). Details about the model development can be found in [71, 83].

a)                b)

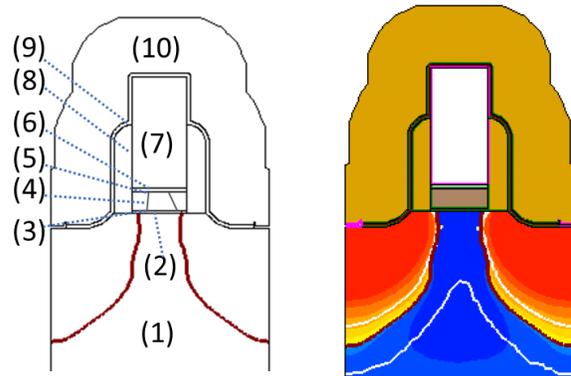

*Figure 13*. 2D geometry of the multi grain FeFET device. a) FeFET geometry consisting of the Si substrate (1), $SiO_2$ interface buffer layer (2), bottom interface layer (3), Si:$HfO_2$ layer (4), top interface layer (5), TiN metal gate (6), poly Si (7), SiN spacer (8), $SiO_2$ liner (9) and nitride layer (10). b) doping profile of the simulated FeFET.

### 4.1 Switching Kinetics-Nucleation Limited Model

The polarization reversal in the ferroelectric films is influenced by various parameters. Among others these are structural and geometric properties of the film as well as electric, thermal and mechanic excitation parameters. For instance, the polycrystalline nature of the film (see section 3 for details), the duration and amplitude of the applied electric field, and the temperature are some of the most important ones. The switching kinetics has both theoretical and practical relevance, because it directly impacts the performance of the device containing the ferroelectric.

Many modeling approaches concerning the switching kinetics have been proposed in order to explain the domain nucleation and growth within the material and to relate these processes to the macroscopic electrical manifestations. Two models are generally accepted in this regard. The Kolmogorov-Avrami-Ishibashi (KAI) model is based on the nucleation and propagation of reversed domains [84]. Here, the coercive voltage $V_C$, obtained as the intersection between the *P-V* hysteretic curve and the voltage axis, follows the simple power-law relationship $V_C \propto f^\beta$, where $f$ is the excitation signal frequency and $\beta$ the fitting parameter. The model has been successfully applied for bulk crystals and clean epitaxial films [85]. On the other hand, polycrystalline disordered ferroelectrics strongly deviate from this relation [86] and therefore, several domain nucleation switching limiting (NLS) models have been proposed [86, 87]. In these NLS models, the time of domain wall movement is neglected compared to the nucleation time. A particularly suitable model for describing field and temperature dependent polarization switching is the one proposed by Du and Chen [87]. Here, the nucleating domains must overcome a critical size defined by an attractive potential. Following a similar approach of Merz [88] for the free energy of domain nucleation within the framework of the classical nucleation theory, a simple expression for the average nucleation waiting time can be found [64]:

$$\tau = \tau_0 \cdot exp\left(\frac{\alpha}{k_B T} \cdot \frac{1}{V^2}\right) \quad (11)$$

where $\tau_0$ is the shortest nucleation time possible, $\alpha$ is a parameter related to intrinsic material properties including domain wall energy and the portion of a switched polarization by the nucleus, $k_B$ is the Boltzmann' constant, $T$ is the temperature, and $V$ is the voltage necessary to switch the polarization.

In order to experimentally study the switching behavior of the ferroelectric hafnium oxide at the domain level, ultra-scaled ferroelectric field-effect transistors (FeFET) were characterized. Due to the ultimately reduced device size, the gate stack contains only one or a few ferroelectric domains, which control the channel conductivity [1, 19]. In this way, it is relatively easy to monitor their switching kinetics by applying proper programming pulses at the gate electrode, and assessing the resulting polarization state by determining the threshold voltage $V_T$ of the transistor. To this purpose, the gate voltage pulse sequence shown in Fig. 14a is adopted. The $V_N$ pulse sets the device in the high-$V_T$ state and with the following $V_P$ pulse an attempt is done to reverse the polarization and set the low-$V_T$ state. Due to the fact that the switching of single domains is a stochastic process when they are excited in the proximity of their coercive voltage, the experiment is repeated 20 times in order to collect the statistical set of data. As a result, the switching time for a fixed $V_P$ spreads over approximately one decade, and its mean value decreases with increasing $V_P$ (Fig. 14b). Reporting the experimental standard deviation $\sigma_{t_{SW}}$ as a function of the average switching time <$t_{SW}$> in a log−log graph (Fig. 14c), a slope of unity over several decades can be clearly identified. This strongly hints at the stochastic switching governed by a Poisson process.

In following the switching in thin films will be modeled and discussed using a purely nucleation-limited approach. Given the small size of grains (the mean grain radius is 10 nm), it is a reasonable assumption that they contain only one domain [89].

We start by supposing that upon the application of the electric field the switching is initiated when a relatively small number of critical nuclei are generated within one grain. The generation rate is given by $\lambda = 1/\tau$ in Eq. (11). Within this picture, the polarization of the grain is considered reversed when a certain critical number $n$ of generated nuclei merge together into a single domain occupying the entire grain. Assuming the nucleation to be the Poisson process, the time elapsing between each critical nucleus generation $\Delta T_i$ will be exponentially distributed as

$$p_{\Delta T_i} = \lambda \, e^{-\lambda \, \Delta T_i} \quad (12)$$

where $p_{\Delta T_i}$ is the probability density function of $\Delta T_i$. Thus the overall domain switching time will be given by the sum of $n$ individual $\Delta T_i$ intervals corresponding to the number of critical nuclei necessary to form the domain:

$$t_{SW} = \sum_{i=1}^{n} \Delta T_i \quad (13)$$

with the mean and variance given by Eqs. (14) and (15), respectively:

$$<t_{SW}> = \frac{n}{\lambda} \quad (14)$$
$$\sigma_{t_{SW}}^2 = \frac{n}{\lambda^2} \quad (15)$$

Now, it is straightforward to determine $n$ and $\lambda$ from the experimental data for $<t_{SW}>$ and $\sigma_{t_{SW}}$ using Eqs. (14) and (15). Generating then $n$ values of exponentially distributed $\Delta T_i$ as dictated by Eq. (12), summing them up according to Eq. (13) and repeating this procedure 20 times for each pulse width ($t_{PW}$) as in the experiment, it is possible to simulate the probabilistic switching. As shown in Fig. 14 d, the switching probability curves, extracted from Figure 14 b, are well fitted across the whole time range, with $n$=5 nuclei for the considered device.

Moreover, the experiment shown in Fig. 14 testifies a clear bias-time trade-off for this nucleation-limited switching [64]. Indeed, referring to Fig. 14b, the device can be programmed at $V_P$ as low as 2.2 V, but only if $t_{PW}$ is longer than 100 μs. However, $t_{PW}$ decreases by more than a factor of 100 when $V_P$ is increased by only 0.6 V. Such exponential dependence, expressed by Eq. (11), represents a source of flexibility and of new opportunities for a future memory design.

a)                                                              b)

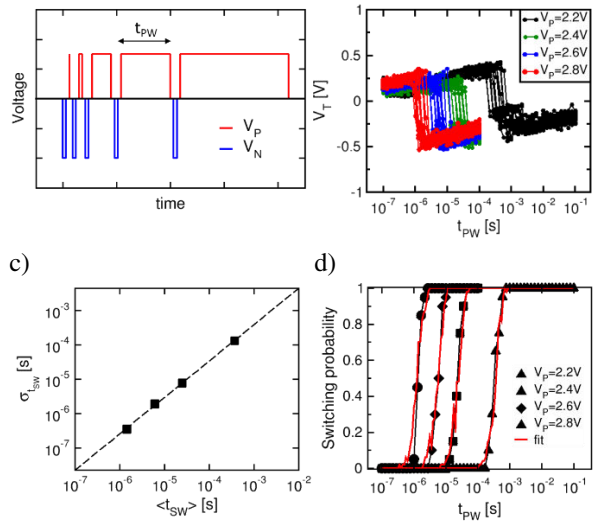

*Figure 14.* Field and time dependence of ferroelectric switching in ultra-scaled FeFETs. a), Gate voltage waveform with logarithmically increasing pulse width $t_{PW}$. After each pulse a fast read out of a transfer curve is carried out. b), $V_T$ vs. $t_{PW}$ graph corresponding to 20 repetitions of procedure in a), shown for four different VP levels. c), Standard deviation vs. mean value of switching time $t_{SW}$ extracted from b). d), Fitting of switching probability curves shown in b).

### 4.2 Single Domain Switching Modeling

As mentioned in the introduction, discovery of the ferroelectric properties in hafnia enabled a tremendous decrease of the memory cell size and high densities of the ferroelectric memories. Reasonably, as a next step beyond further geometrical scaling, multi-level-cell operation could result in much increased memory densities per chip area. In 2015 in ref. [63] a single-domain switching in ultra-scaled Si:HfO$_2$ based FeFET which results in a stepwise change in the $V_T$-characteristics for both program and erase sequence (Figure 15) was reported. To study the switching to the low $V_T$ state the measurement was carried out as following: after initial 100 bipolar cycles used for device preconditioning and reach of the fully woken-up state, a reference ERS pulse defining the high reference $V_T$ was applied. After the read-out of the reference $V_T$, the PRG pulse was incrementally increased in 50 mV steps. It should be noted that every ERS and PRG operation was followed by a read-out. Vice versa to PRG sweep, pulse sequence was applied for the ERS sweep (red characteristics of the Figure 15b). Considering an average grain diameter in the order of 10 nm, these ultra-scaled devices (gate length = 28 nm and width = 80 nm) are expected to contain about two to four ferroelectric grains. Further scaling of the device while preserving the grain/domain size would result in increased variability of the MW and hence might influence the behavior of the whole array. However, different technological measures can be adopted to counteract that issue by decreasing grain size or increasing the uniformity of the ferroelectric film properties.

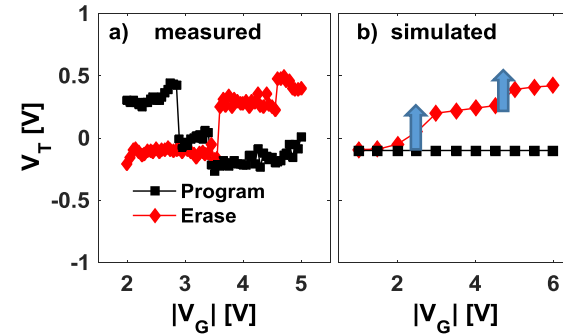

*Figure 15.* a) Discrete switching events for both PRG (black) and ERS (red) of the cell. b) Simulation of the discrete switching based on the 2D multi-grain TCAD model of FeFET.

Here, each discrete switching event which abruptly shifts the $V_T$ is represented with a discrete step in $V_T$ characteristics. In the same study [63] a simulation was performed to verify the model and physical process behind discrete switching events. Details of the model were reported in [71]. A three-grain 2D TCAD model (see Fig. 13) was utilized for the simulation of the discrete switching. Each grain is characterized by a different $E_C$ value (1 MV/cm; 1.5 MV/cm and 2 MV/cm). The simulation results for the 2D multi-grain TCAD model and its comparison with measurements are shown in Figure 15. Similar to the measured characteristics, steps in simulated $V_T$-$V_G$ dependence are visible. However, in contrast to the measured abrupt switching steps, the simulated device exhibited smeared out switching steps. Such behavior is a consequence of the nature of the Preisach based model provided in TCAD as discussed in Section 3. In contrast to Landau-Khalatnikov model characterized with abrupt switching, the Preisach-model which sums up discrete square-shaped hysterons and averages it over the parameter distribution (e.g. internal bias fields and coercive fields). It should be noted that even within the hysteresis of the three single grains, partial switching and sub-loops are allowed, consequently yielding a gradual shift of the $V_T$. Further, it can be debated if this is physically justified and how small a single hysteron is. It could represent a single unit cell or also a single grain. The results found in this studies, suggest that the latter assumption is more appropriate. Therefore, for ultimately scaled devices comprising just a few grains, the usage of a Preisach-model that accounts for a large ensemble of domains is no longer suitable. Instead, it can be concluded that for the purposes of the simulation of discrete switching events and addressing of the single domain switching kinetics a Landau-Khalatnikov (L-K, see Section 4.3 for details) model for emulation of ferroelectrics is necessary. Detail about the L-K model and its advantages will be addressed in following section.

### 4.3 Landau-Khalatnikov Model

Starting purely from symmetry considerations, Landau theory is a powerful tool for describing a multitude of phase transition phenomena based on thermodynamic free energy potentials [90]. This phenomenological approach was first applied by Devonshire to model phase transitions in ferroelectric materials [91]. Close to the transition temperature $T_C$, a high symmetry paraelectric phase transforms into a lower symmetry ferroelectric phase. The simplest way to describe this phase transition is by doing a series expansion of the free energy $u$ using the order parameter $P$, which is the electrical polarization:

$$u = \alpha P^2 + \beta P^4 + \gamma P^6 - EP. \quad (16)$$

Here, α, β, γ are the ferroelectric anisotropy constants and $E$ is the electric field in the ferroelectric. Below $T_C$, for second-order phase transitions, α < 0, β > 0 and γ = 0, whereas for first-order phase transitions α < 0, β < 0 and γ > 0. Fig. 16a shows the free energy as a function of $P$ for different applied fields $E$ for a second-order transition. Without an applied field there are two degenerate energy minima corresponding to the remanent polarization states (+$P_r$ and -$P_r$). When a field $E$ is applied, one minimum is becoming energetically more favorable while the other one increases in energy and ultimately vanishes at the coercive field $E_C$. By differentiating Eq. (16) with respect to $P$ and setting d$u$/d$P$ = 0, we obtain the S-shaped $P$-$E$ relationship corresponding to the polarization hysteresis as shown in Fig. 16b. It should be noted here, that in the case shown in Fig. 16b the switching at $E = E_C$ would be instantaneous, which is obviously not physical.

Therefore, to calculate more realistic switching dynamics based on the free energy in Eq. (16), the Landau-Khalatnikov (LK) equation is applied [92]:

$$\rho \frac{\partial P}{\partial t} = -\frac{\partial u}{\partial P}, \quad (17)$$

where ρ is a damping constant of dimension [Ω·m], which corresponds to the internal resistivity of the ferroelectric which limits the change of $P$ with time. In case of Fig. 16b, ρ would be zero.

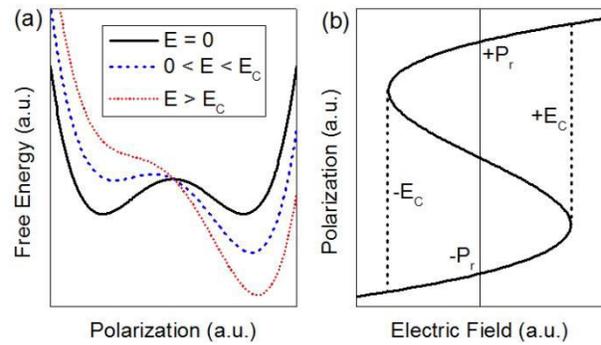

*Figure 16.* a) Landau free energy potential as a function of the applied field $E$. b) Polarization-electric field hysteresis calculated from a).

It should be noted, that this is a mean-field theory which assumes homogeneous properties throughout the whole ferroelectric. While this is obviously not true for polycrystalline thin films like $HfO_2$, we will show how inhomogeneities can be incorporated into the Landau-Devonshire model. A general extension of the model to include a gradient term $(\nabla P)^2$ was proposed by Ginzburg and Landau [93], which is also used in more complex ferroelectric phase field models [94]. However, we will propose a simpler approach to describe $HfO_2$ based ferroelectrics, which is much less computation intensive than phase field models and much more accurate than homogeneous Landau theory.

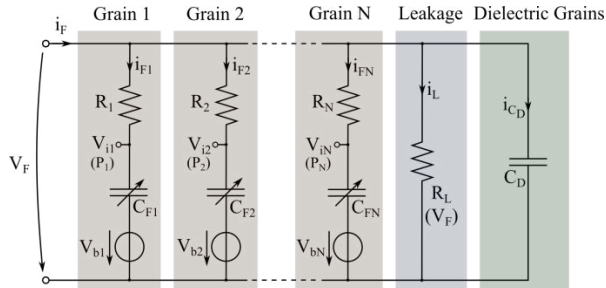

*Figure 17.* Equivalent circuit representation of the multigrain Landau-Khalatnikov model for a ferroelectric HfO$_2$ capacitor.

Generally, LK models based on homogeneous Landau-Devonshire theory are not suited to describe polarization switching in ferroelectric thin films, since local effects like domain nucleation and growth are known to govern polarization reversal [95]. Therefore, a modified LK approach was proposed [96], where the ferroelectric film is partitioned into smaller regions, which are electrically in parallel and are each described by Eq. (16) and (17). By introducing spatial distributions of the parameters α, β, γ and ρ, non-uniformity of the film can be modeled [96].

In the case of HfO$_2$ based ferroelectrics, the polycrystallinity and grain structure is especially of interest. Some grains might be paraelectric and therefore can be modeled as a regular linear capacitance in parallel. Each ferroelectric grain, on the other hand, is described by the LK equation. Then, in case of a ferroelectric capacitor with two metal electrodes, the total charge on the capacitor is calculated as

$$Q_{FE} = A_{FE}\left(\varepsilon_0 E + \frac{1}{N}\sum_{j=1}^{N} P_j\right) \approx \frac{A_{FE}}{N}\sum_{j=1}^{N} P_j, \quad (18)$$

where $A_{FE}$ is the area of the capacitor plates, $\varepsilon_0$ is the vacuum permittivity, $N$ is the total number of grains and $P_j$ is the polarization of the grain $j$. It is assumed that all grains have the same area $A_{FE}/N$ and that each grain has a homogeneous polarization. While this is an approximation, it seems reasonable since grain sizes in HfO$_2$ based ferroelectrics are typically in the same order as the film thickness $T_{FE}$ (5-30 nm) [97]. Additionally, grain boundaries generally run from one electrode to the other, perpendicular to the film and nucleation limited switching kinetics show (see section 4.1) that individual parts of the film switch independent of each other [64]. Therefore, the assumption of modeling non-interacting grains electrically in parallel seems justified. The equivalent circuit representation of this model is shown in Fig. 17. Here, $R_j = \rho_j T_{FE}/A_{FE}$ is the internal resistance, $V_{ij} = 2T_{FE}(\alpha_j P_j + 2\beta_j P_j^3) + V_{bj}$ is the internal voltage, $V_{bj}$ is the internal bias voltage [98] and $i_{Fj}$ is the current of grain $j$. $C_{Fj}$ corresponds to the non-linear capacitance of each grain ferroelectric grain which is given by $(d^2u/dP^2)^{-1}A_{FE}/T_{FE}$. The leakage current $i_L$ is modeled by a parallel resistance $R_L$ which is a function of $V_F$ (typically an exponential dependence). All non-ferroelectric grains can be combined into a parallel capacitance $C_D$. Using this multigrain LK approach, experimental data from standard P-V hysteresis measurements can be fitted by adjusting the α, β, γ and ρ parameter distributions [96], as shown in Fig. 18 for a 18 nm thin ferroelectric Gd:HfO$_2$ film, with N = 10,000.

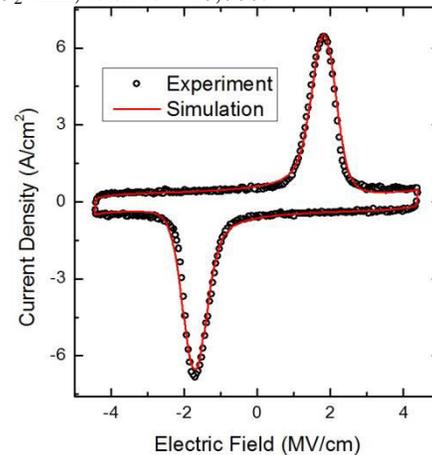

*Figure 18.* Comparison of experimental data and multigrain Landau-Khalatnikov simulation of an 18 nm thin Gd:HfO$_2$ capacitor. Current response to a triangular voltage signal with a frequency of 10 kHz.

While this example shows the very good agreement of the multigrain LK model with data for rather large capacitors ($A_{FE}$ = 33,000 µm$^2$), it should still result in a more accurate representation of HfO$_2$ based ferroelectrics in nanoscale devices. For instance, it should be possible to describe the single-domain switching seen in FeFETs scaled to the 28 nm node [64] more accurately compared to the Preisach model (see section 4.2 for details). However, new complications arise when simulating a FeFET structure due to different boundary conditions compared to a capacitor: When the ferroelectric HfO$_2$ is deposited directly onto the Si channel a SiO$_2$ interface forms [64]. Therefore, the potential across each grain will be locally different and a particular spatial distribution of grains will have to be assumed. In case of a regular capacitor with metal electrodes, the actual spatial position of each grain is irrelevant.

Additionally, completely new physical phenomena arise when using the LK model compared to other ferroelectric models. Based on Landau theory, it was proposed that the region of negative slope of the S-shaped *P-E* curve (corresponding to a negative capacitance (NC)) could be used to build a FET with subthreshold swing lower than 60 mV/decade at room temperature [99]. Such NCFETs based on $Hf_{1-x}Zr_xO_2$ have already been demonstrated experimentally [100] and are a promising way to further reduce the supply voltage and thus the power dissipation in future logic circuits. However, all device simulations reported so far applied a homogeneous single domain Landau model [101], which cannot accurately predict effects of non-uniformity in $HfO_2$ based ferroelectrics. The device structure of an NCFET is very similar to a regular FeFET, since also a ferroelectric is used in the gate stack, just with different desired parameters ($T_{FE}$, $P_r$ and $E_C$). Therefore, using an LK based approach would be beneficial to combine existing models for both memory FeFETs and logic NCFETs into a single comprehensive modeling framework.

## 5 Mixed-mode Modeling and Circuit Model

After the single device simulations discussed up to now, next level in simulation hierarchy is achieved through the grouping of these devices into circuits and performing the mixed-mode simulations. These mixed-mode TCAD simulations extend the single device model not only by adding different device models but also by combining it with the HSPICE compact models. Based on that it is not hard to conclude that these simulations are positioned hierarchy-wise between the device simulations and compact models. In the first part of this chapter a mixed-mode simulation of the AND array architecture of FeFET will be reviewed, whereas the second part is devoted to a circuit model of the FeFET.

### 5.1 TCAD Simulation of Elementary Array Architectures

When looking at semiconductor memories the first step to the circuit level requires array-level simulations. This type of simulation is required in order to verify the functionality of each individual memory cell when embedded into a larger array of multiple cells. For example write disturb is one of the parasitic effects occurring solely when a single cell is incorporated in a certain memory architecture. In order to illustrate this, the so called AND architecture is chosen as an example (see Fig. 19).

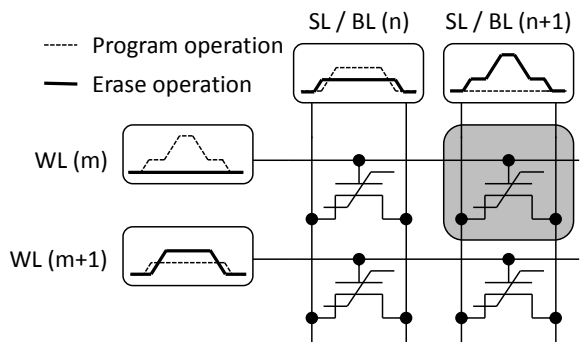

*Figure 19*. Program and erase operation of FeFET cells in AND architecture unit cell representation. The transient voltage trains on the terminals are illustrated for the program (dashed lines) and the erase (solid lines) operation. The cell to be written is highlighted in grey.

For FeFET cells arranged in an AND configuration, a general way to write a selected cell is given by the so called $V_{DD}/3$ scheme described elsewhere [102, 103]. This operational scheme allows for reducing the write disturb occurring in one-transistor (1T) FeFET array architectures. That is, when a cell is selected to be written (i.e. either for program or erase operation), the shared terminals called wordline (WL), bitline (BL) and sourceline (SL) have to be raised or lowered to certain write and inhibit voltages. At the end, the voltages have to be chosen such that the voltage drop across the gate stack of the selected cell induces a polarization reversal in the ferroelectric film while at the same time all neighboring cells must not change their binary state. The $V_{DD}/3$ scheme in principal allows for a complete voltage drop of $V_{DD}$ across the selected cell whereas the unselected cells only experience one third of this write voltage as parasitic disturb.

In order to verify the functionality of such an operational scheme for a specific FeFET device geometry (see Fig. 13 in section 4), TCAD simulation can be utilized. For the given cell array unit cell of Fig. 19, various worst case scenarios have to be considered (see also [103] for a detailed description). The results of the respective TCAD simulation are shown in Fig. 20.

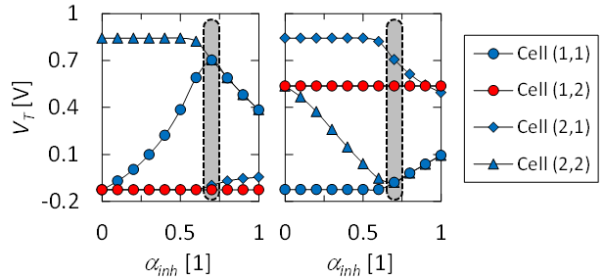

*Figure 20*. Verification of $V_{DD}/3$ operational scheme by TCAD simulation. The evolution of threshold voltages $V_T$ over the inhibit factor $\alpha_{inh}$ is shown for all cells of the AND architecture unit cell representation. The indices of the cells in the legend encode rows / wordlines and columns / bitlines and sourcelines. Cell (1,2) is the selected cell. a) represents the program operation and b) the erase operation performed on cell (1,2).

The TCAD simulations where performed such that all worst case disturb scenarios on the respective cells in the AND architecture unit cell representation were analyzed as a function of the so called inhibit factor $\alpha_{inh}$. With respect to Fig. 19, the inhibit factor determined the voltages applied to the respective signal lines:

Program operation:

$$V_{WL(m)} = V_P \quad (19)$$

$$V_{WL(m+1)} = \frac{\alpha_{inh}}{2} \cdot V_P \quad (20)$$

$$V_{S/BL(n)} = \alpha_{inh} \cdot V_P \quad (21)$$

$$V_{S/BL(n+1)} = 0 \quad (22)$$

Erase operation:

$$V_{WL(m)} = 0 \quad (23)$$

$$V_{WL(m+1)} = \alpha_{inh} \cdot V_P \quad (24)$$

$$V_{S/BL(n)} = \frac{\alpha_{inh}}{2} \cdot V_P \quad (25)$$

$$V_{S/BL(n+1)} = V_P \quad (26)$$

Accordingly, for the program operation in Fig. 20 on the left-hand side it can be observed that only for an inhibit factor of around 0.7, it is possible to reduce the worst case disturb for cell (1,1) appropriately. To be more specific, the worst case for cell (1,1), when cell (1,2) shall be programmed into the low-$V_T$ state, is present when cell (1,1) originally resides in a high-$V_T$ state. This state can only be maintained when the unselected column n is raised to around 0.7 times the write voltage $V_P$. This verifies that the $V_{DD}/3$ scheme is indeed best suited for minimizing disturb.

The same analysis can be performed for an erase operation on cell (1,2) via the so called Positive-Source-Drain-Erase-Scheme (PSDES) [103]. According to Fig. 20, also there the $V_{DD}/3$ scheme, i.e. at an inhibit factor of around 0.7, works best for reducing write disturb, in this case erase disturb in particular. Especially with respect to the PSDES scheme it is interesting to note that the high-$V_T$ state after the erase operation on cell (1,2) is not as high as it is for cell (2,1) which had previously been written into the reference state by a so called negative gate erase (NGE) operation. This is rooted in the effectiveness of the PSDES and a detailed computational study can be found in [83]. After a in-depth discussion of the single device behavior and influence of the different cells on the neighboring cells in an array, a circuit model will be discussed next.

## 5.2 Circuit model of ferroelectric devices

Different circuit models were proposed in the literature to model ferroelectric materials in electron devices. Prior to the actual fabrication step of devices they allow the evaluation of designed circuits. Simulation models comprise mathematics-based approaches where the hysteresis loop is reproduced by mathematical functions, switching current models, the parallel elements method as well as physics-based models (see section 3) based on nucleation mechanisms (KAI and NLS model; see section 4.1) or the double-well potential of the Gibbs free energy (Landau-Khalatnikov model; see section 4.3).

The mathematical approach includes the Preisach model [104, 105, 106, 107, 108, 109, 110] already briefly introduced in section 3. Statistically distributed dipoles with individual coercive fields and a specific density function with respect to the coercive fields are superimposed to obtain the overall polarization. If the density function is Gaussian type, the resulting polarization becomes a tanh-function. In the given example, the full hysteresis loop is reproduced by two tanh-functions which are shifted along the E-axis by $\pm E_C$ (see Eq. (4) for details). Another modelling approach relates to the polarization switching current of the ferroelectric [111, 112, 113, 114]. Here the current is modelled based on the rate equations of the switching dipoles [114], pre-calculated polarization current [113] and first order relaxation processes [111, 112]. On the other side the parallel elements method [115, 116] emulates the structure of a ferroelectric material as a parallel circuit of several ferroelectric domains. Single domains are simulated by a Schmitt trigger [Sun05, Yamamoto] and voltage controlled capacitor and resistor [115, 116]. Finally physics-based models are based on the LK-equation [117, 118, 119, 121] or the KAI-model [120]. The non-linearity of the ferroelectric is either reproduced abstractly by controlled current and voltage sources [117] or, more circuitry-wise, by a series connection of a non-linear capacitor $C_N$ and a resistor $R_N$ with a linear capacitor $C_0$ in parallel [118, 120, 119]. The linear capacitor originates from the consideration of the total charge in the ferroelectric.

All previously-mentioned models are suitable to reproduce the hysteretic behavior of the ferroelectric. Switching current models can reproduce transient switching and DC behavior, but the switching current has to be known in the first place or assumptions about its behavior have to be made. The parallel elements method, though the single dipoles can be modeled in a very simple way (e.g. Schmitt trigger), can lead to increasing simulation times because of the numerous parallel branches. The same approach accounts for the physics-based models when several domains are modelled. The latter ones are favorable when the priority is given to a behavioral model which emulates the physics of the switching kinetics. Static and dynamic behavior can be described and only a low number of parameters is needed to describe a single dipole.

The Model of the ferroelectric material described in the following (Figure 21) is based on the approach of Aziz et al.[118] who made use of the LK-equation (Eq. (16)-(17); section 4.3.) to express the voltage drop $V_{FE}$ across the ferroelectric as:

$$V_{FE} = T_{FE}\left(\frac{\rho}{A_{FE}}\frac{dQ_P}{dt} + \frac{\alpha Q_P}{A_{FE}} + \frac{\beta Q_P^3}{A_{FE}^3} + \frac{\gamma Q_P^5}{A_{FE}^5}\right) \quad (27)$$

where α, β, γ and ρ are the parameters of the LK-equation introduced in section 4.3. $T_{FE}$ and $A_{FE}$ are the thickness and the area of the ferroelectric, respectively.

The first term corresponds to a series resistor with the resistance $R_N = T_{FE} \cdot \rho / A_{FE}$ and the following terms reflect the non-linearity of the ferroelectric capacitor. This non-linearity (of the ferroelectric) is emulated by a voltage controlled voltage source $V_N$ which is dependent on the charge $Q_P$. To get $Q_P$, the current $I_{VN}$ through the voltage source is integrated over time by an integrator circuit. Based on this simulation model (Figure 21) clamped on standard BSIM transistor model (Figure 22a) typical on-/ off-states of a FeFET are observed (Fig. 22b).

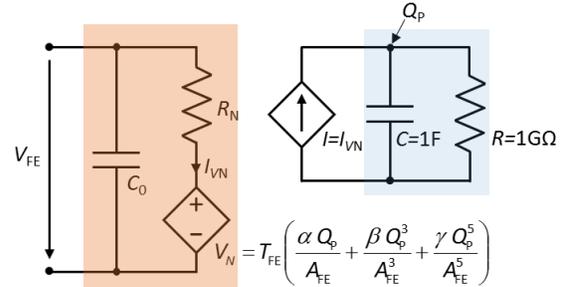

*Figure 21.* Circuit model of a ferroelectric capacitor based on the Landau-Khalatnikov equation. Ferroelectric capacitor and integrating circuit are marked with orange and blue rectangle, respectively.

In contrast to the model described here, previously discussed model (see section 4.3) apart from multigrain implementation, includes internal bias fields and nonlinear capacitance, whereas the model reported by Aziz et al. [118] emulates this nonlinearity with voltage source. Including parallelization (to emulate multi-domain nature of the devices) and additional constant voltage source (to emulate internal bias field present) a direct conversion from model described here to model discussed in section 4.3 is possible.

To assess the realistic device performance trapping effects have to be considered and incorporated. Trapping of electrons and holes as a parasitic effect compensates the MW of the device by shifting the threshold voltage towards more positive and negative values, respectively. This parasitic threshold voltage shift is proportional to number of defects filled for certain applied pulse amplitude and length. A straight forward solution to address this parasitic behavior would be to account this shift by an auxiliary sub-circuit (see Fig.22a; charge trapping emulation), similar to one reported in [122]. Other implementations like adding a second controlled voltage source in series are possible. However, since the charge trapping and de-trapping effects occur mainly in the interface oxide region and thus depend on the respective electrical fields, the additional circuit should be placed in-between the transistor model and the voltage source modeling the ferroelectric polarization effect.

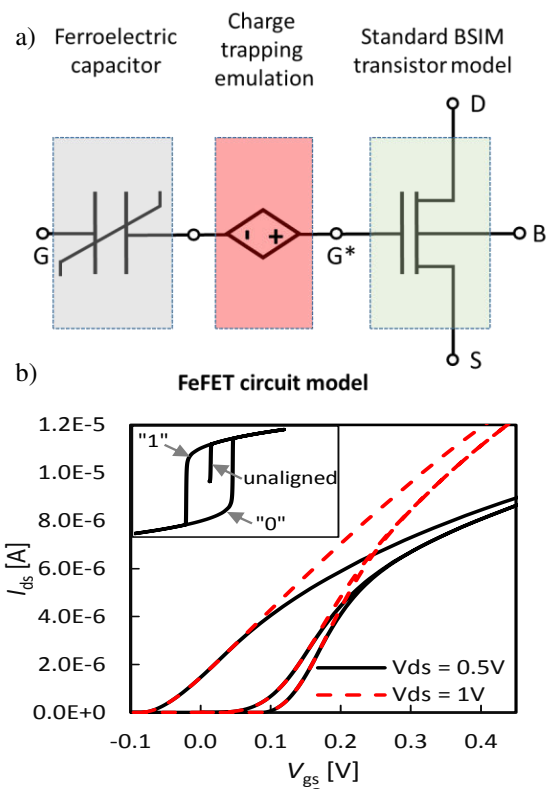

*Figure 22.* a) Circuit model of FeFET comprising a FE capacitor, controlled voltage source (charge trapping emulation) and standard BSIM model of FET transistor. b) Results for the on- and off-state of the FeFET from the circuit model shown in fig. 21. for $V_{ds}$ = 0.5 V (yellow, solid line) and $V_{ds}$ = 1 V (orange, dashed line). The middle curve shows the initial curve (dipoles unaligned). The high $V_T$-state corresponds to polarization state "0", the low $V_T$-state corresponds to polarization state "1" (see inset).

## 6 Summary and Outlook

The discovery of ferroelectric properties within $HfO_2$ revived the interest in ferroelectric memories leading to a rapid development and impressive device scaling in the last 10 years. In this article, we provided an overview of modeling hierarchies and approaches which were developed in this timeframe not only to unveil the origin of the ferroelectricity in binary oxides and physical mechanism taking place with field stressing of the device, but also as strategies that have to be considered when modeling and designing the future ultimately scaled devices for practical applications. Even though there is a number of challenges to be overcome, both in the experimental realization and the modelling framework, ferroelectric memories based on hafnium oxide show a huge potential for becoming a wide-spread, high-speed and low-power non-volatile memory solution of the future.